\documentclass[12pt]{article}
\usepackage{geometry}
\title{Evolution of Complexity}
\author{Carlos Gershenson$^{1,2}$ and Tom Lenaerts$^{3,4}$\\
\mbox{}
$^1$ CLEA, Vrije Universiteit Brussel, Brussels, Belgium\\
\mbox{}
$^2$ New England Complex Systems Institute, Cambridge, MA, USA\\
\mbox{}
$^3$ SWITCH, VIB, Brussels, Belgium\\
\mbox{}
$^4$ Vrije Universiteit Brussel, Brussels, Belgium\\
$\{$ cgershen $|$  tlenaert$\}$ @vub.ac.be}

\begin{document}
\maketitle


The evolution of complexity has been a central theme for Biology \cite{Bonner1988} and Artificial Life research \cite{problems}. It is generally agreed that complexity has increased in our universe, giving way to life, multi-cellularity, societies, and systems of higher complexities. However, the mechanisms behind the complexification and its relation to evolution are not well understood.  Moreover complexification can be used to mean different things in different contexts. For example, complexification has been interpreted as a process of diversification between evolving units \cite{Bonner1988} or as a scaling process related to the idea of transitions between different levels of complexity \cite{szathmary97}.  Understanding the difference or overlap between the mechanisms involved in both situations is mandatory to create acceptable synthetic models of the process, as is required in Artificial Life research.  

Concretely, many open questions related to the evolution of complexity can be asked. Some were proposed in the call for papers, of which some are addressed in this issue:
\begin{enumerate}
\item How could complexity growth be measured or operationalised in natural and artificial living systems?
\item How can existing data from nature be brought to bear on the study of this issue?
\item What are the main hypotheses about complexity growth that can actually be tested today?
\item Are the principles of natural selection as they are currently understood sufficient to explain the evolution of complexity in living systems?
\item What are the environmental and other constraints of the evolution of complexity in living systems?
\item What is the role of developmental mechanisms in the evolution of complexity in living systems?
\item What conditions could reduce evolved complexity in living systems?
\item How factors allow the evolution of complexity in living systems to be manipulated and controlled?
\item What models are most appropriate for understanding the evolution of complexity in living systems?
\end{enumerate}

To clarify the understanding in the ALife community, some workshops have been organized dedicated to this topic \cite{HeylighenEtAl1999,wdh02a}, including our workshop at the ALife X conference \cite{GershensonLenaerts2006}, held at Indiana University in Bloomington.  Moreover, overlapping discussions have been published in another special issue of the Artificial Life journal \cite{specis05}. As the consequence of the interest and creative discussions which surrounded our workshop, it was decided to organize another special issue to collect the most interesting contributions. We received an enthusiastic response from the community, reflected in a large number of submissions. From these, ten articles were selected for their relevance, clarity of presentation, quality, originality, and contribution to ongoing debates. That is to say, some papers may be controversial, but we decided to include them to stir healthy discussions.  The articles are concerned with ways to generate, measure and formalize complexity.  Moreover, due to the long history concerning this topic, articles provide meaningful pointers to earlier work.  Given all this information we hope that this special issue will provide the incentive for other Artificial Life researchers to contribute to the search for answers to the previous questions, or to provide a better definition of the questions themselves.   In summary, the following issues are addressed:

Olof G\"{o}rnerup and James Crutchfield study hierarchical organization in the ``finitary process soup" model, where global complexity is due to the emergence of successively higher levels of organization. This adresses the question of how a relatively small number of genes can code for the high complexities of different organisms.

The Avida system is used by Charles Ofria, Wei Huang and Eric Torng to study the gradual evolution of complexity in digital organisms, which leads to the sudden emergence of complex features.

Philip Gerlee and Torbj\"orn Lundh also use the Avida system, but to study the structure of the genetic architecture of digital organisms. They find that the gene overlap follows a scale-free distribution, whose slope depends on the mutation rate, which is also related to a more complex genetic architecture.

Nicholas Geard and Janet Wiles present \emph{LinMap}, an interactive visualization tool for exploring the structure of complexity gradients in evolutionary landscapes. Interestingly, different definitions of complexity produce different gradients across the same landscape.

Attila Egri-Nagy and Chrystopher L. Nehaniv go beyond complexity measures, using hierarchical decomposition to assess structural complexity of systems formally modelled as automata. This decomposition allows a better grasp of the complexity of organisms.

Dominique Chu adresses the problem of the notion of complexity, since there is no consensus on what complexity is. Chu recommends abandoning attempts to formalize complexity for the time being, and use complexity informally in research, which might eventually lead to a more formal notion.
 
A broad review of literature on the evolution of complexity is offered by Thomas Miconi, followed by a discussion on whether the ``arrow of complexity" is a passive or active trend, and the role of evolution in the creation of this trend. Certain conditions for the ``arrow of complexity" are also proposed.

Alexander Riegler argues for extending the notion of ``natural selection" to include also ``internal selection" in order to account for the evolution of complex systems, defending that canalization is an indispensable ingredient for evolutionary processes.

Stanley Salthe adresses whether  the principles of natural selection as they are currently understood sufficient to explain the evolution of complexity in living systems, presenting a critique of natural selection theory, arguing that it cannot, by itself, account for increases in complexity.

George Kampis and L\'aszl\'o Guly\'as retake this question and several others, arguing for the importance of a dynamically defined, deeply structured and plasticity-bound phenotype in the evolution of complexity.

Overall, the authors were preoccupied with appropriate definitions of complexity. It seems that there is no ``best" definition independent of a context. However, some definitions of complexity are proving their usefulness. Probably we do not need a single definition, as different definitions are useful in different contexts. Still, our understanding of complexity continues to increase.

Another important topic discussed by the authors was evolutionary theory, which has been evolving since Darwin's times. The papers suggest that complexity plays a key role in evolution. Having a better understanding of complexity will enable us to broaden our knowledge of evolution.

This special issue contains important contributions, but also further obstacles.
We are far from answering all of the questions presented above. Still, we hope that the articles will foster further discussion and lines of research. As Dominique Chu notes, we do not need an agreed definition of complexity to study its evolution. We are making progress, as the notion of ``complexity" continues to evolve. This special issue is just a step in the evolution of ``evolution of complexity" studies.

\section*{Acknowledgements}

We would like to thank all the program committee members: Chris Adami, Lee Altenberg, Mark Bedau, Hugues Bersini, John Bonner, Dominique Chu, Jim Crutchfield, Bruce Edmonds, Carlos Gershenson, Mario Giacobini, Francis Heylighen, Tom Lenaerts, Juan Juli\'{a}n Merelo Guerv\'os, Barry McMullin, Chrystopher Nehaniv, Jorge Pacheco, Tom Ray, Jon Rowe, Stanley Salthe, Cosma Shalizi,  Richard Watson, and Larry Yaeger for taking time to provide in depth reviews and suggestions.  We thank all researchers who sent articles for this special issue. We greatly appreciate the support from Mark Bedau as Editor of Artificial Life.

We thank Mark Woolsey for proof-reading the manuscript. We acknowledge support by the FWO and the New England Complex Systems Institute.

\bibliography{ECO-ALJ-Intro}
\bibliographystyle{alj}

\end{document}